\documentclass[12pt]{article}
\usepackage{epsfig}
\usepackage{a4}

\def\to{\rightarrow}

\newcommand\nue{{\nu_e}}
\newcommand\anue{\bar{\nu}_e}
\def\sq2{sin^2(2\Theta)}

\def\NOE{{\em N\raise.5ex\hbox{O \kern-0.47em}E\kern.4em}}

\def\FLUKA{{\sc FLUKA}}

\def\dms2{\Delta m^2}
\def\Dm32{\Delta m^2_{32}}
\def\Dm12{\Delta m_{12}^2}
\def\tet12{\theta_{12}}
\def\tet23{\theta_{23}}

\def\etal{{\it et\ al.}}

\def\PLB{{\em Phys. Lett.}  B}
\def\PRL{{\em Phys. Rev. Lett.}}

\def\PRD{{\em Phys. Rev.} D}

\begin{document}

\thispagestyle{empty}
\begin{flushright}
{\tt ICARUS-TM/03-02}\\ 
\today
\end{flushright}
\vspace*{1cm}
\begin{center}
{\Large{\bf Supernova Neutrino Detection \\ in a  liquid
Argon TPC}}\\
\vspace{.5cm}
{\large A. Bueno}\footnote{Antonio.Bueno@cern.ch}

Dpto. de F\'{\i}sica Te\'orica y del Cosmos and CAFPE, \\
Universidad de Granada, Spain

\vspace*{0.3cm}
{\large I. Gil-Botella}\footnote{Ines.Gil.Botella@cern.ch},
{\large A. Rubbia}\footnote{Andre.Rubbia@cern.ch}

Institut f\"{u}r Teilchenphysik, ETHZ, CH-8093 Z\"{u}rich,
Switzerland
\end{center}
\vspace{2.cm}
\begin{abstract}
\noindent
A liquid Argon TPC (ICARUS-like) has the ability to detect neutrino bursts from 
type-II supernova
collapses, via three processes: elastic scattering by electrons from all
neutrino species, and $\nu_e$ charged current absorption on $Ar$ with production
of excited $K$
and $\bar\nu_e$ charged current  
absorption on $Ar$ with production of excited $Cl$.
In this paper we have used
new calculations in the
random phase approximation (RPA) for the absorption processes. 
The anti-neutrino reaction
$\bar\nu_e \ ^{40}Ar$ absorption has been considered for the
first time. For definiteness, we
compute rates for the 3 kton ICARUS detector:
244 events are expected from a supernova at a distance of
10 kpc, without considering oscillation effects.
The impact of oscillations in the neutrino rates will be discussed in a
separate paper. We also discuss the ability to determine
the direction of the supernova.
\end{abstract}

\newpage
\pagestyle{plain} 
\setcounter{page}{1}
\setcounter{footnote}{0}

\section{Introduction}
\label{sec:one}

When a massive star (eight or more times the mass of the Sun) runs out
of nuclear fuel, it develops a layered structure, which heavier elements
lay in deeper layers, the innermost element being iron. This
configuration becomes eventually unstable and leads to a catastrophic
collapse of the star core in a time window of a few thousandths of a
second, since the star cannot resist the pressure of its internal
gravitational force. This collapse leads to an explosion that is known
as type II supernova ~\cite{bahcall}. The expected rate for such an
event to occur in our Galaxy is one every thirty years~\cite{snrate}. 

The binding energy released during this process is enormous: $10^{53}$
ergs. The high density prevents the emission of
photons, therefore $99\%$ of the energy comes out in the form
of neutrinos of all types. A small fraction of them, about $1\%$, 
are produced during the neutronization process through the reaction 
$e^- + p \to \nu_e + n$, while the rest are $\nu\bar\nu$ pairs from
later cooling reactions.

On their way out, neutrinos interact inside the stellar core 
which is an extremely dense and neutron-rich medium. The different
scattering reactions, underwent by every neutrino species, result in
different energy distributions for the emitted neutrinos. Muon
and tau neutrinos interact only through weak neutral currents, thus they
decouple deep inside the core, where a higher temperature exists, and
therefore are emitted with higher energies compared to $\nu_e$'s and
$\bar\nu_e$'s. In addition, $\bar\nu_e$'s suffer fewer charged
current interactions than $\nu_e$'s, so they decouple in hotter regions,
thus having an energy spectrum that is harder than that of $\nu_e$'s. 

The neutrino signal from a supernova 
rises first steeply and then falls exponentially with
time (see \cite{burrows}). The experimental detection of a neutrino
signal from a supernova is therefore expected to occur in a time window of about
10 seconds. This implies that detector thresholds can, in principle, be 
set quite low since backgrounds are highly suppressed given the narrow 
time window in which the burst takes place.

The detection of neutrinos emitted by SN1987A in
Kamiokande~\cite{kamioka} and IMB~\cite{imb} began the era of neutrino
astronomy. Nowadays, 
several running or planned experiments
(SuperKamiokande, SNO, LVD, KamLAND, Borexino...) have the
capability to unambiguously detect supernova
neutrinos~\cite{Scholberg:2000ps,Cadonati:2000kq}. 

As discussed in~\cite{bahcall,Cennini:1993tx}, a liquid Argon
TPC like in ICARUS has the ability 
to detect neutrino bursts from supernova 
collapses. Preliminary estimates of the expected rates
yielded 76 events (44 from $\nu_e$ absorption and 32 
from electron scattering assuming a 5 MeV electron detection
threshold)\cite{Cennini:1993tx} for a supernova at 10~kpc. 
For the absorption cross section, a threshold of 11 MeV was
taken into account\footnote{in ref.~\cite{Raghavan:1986hv} a threshold 
$Q = 11.5$ is quoted in units of $m_e$.}. The energy threshold
for this Fermi transition is however 5.885 MeV~\cite{Raghavan:1986hv}. In addition, 
the Gamow-Teller transitions were not included in the 
total absorption rate. 

In this paper, we improve the estimates for supernova neutrino rates
taking into account latest developments: we use
new calculations in the
random phase approximation (RPA) for the absorption processes\cite{martinez}
 and
consider all the processes that contribute to the 
total number of expected neutrino events, including for the first time
the absorption reaction $\bar\nu_e \ ^{40}Ar$. We discuss the ability
to determine the direction of the supernova.

\section{Supernova neutrino signatures}
\label{sec:reso}

Models of type II supernovae predict that neutrinos are emitted with
a thermal spectrum, with a temperature hierarchy among neutrino
flavours: $T_{\nu_e} < T_{\bar\nu_e} <T_{\nu_\mu,\nu_\tau,\bar\nu_\mu,\bar\nu_\tau}$. 
The neutrino energy spectra can be described by a
Fermi-Dirac distribution~\cite{Langanke:1996he}:
\begin{equation}
\frac{dN}{dE_\nu} = \frac{C}{T^3}\frac{E_\nu^2}{1 + e^{(E_\nu/T-\eta)}} N_\nu
\end{equation}
where $C=0.55$; $T$ is the temperature (MeV); $E_\nu$ 
the neutrino energy (MeV); $\eta$ the chemical potential and
$N_\nu$ the number of expected neutrinos of a given
species. We use $\eta=0$~\cite{Langanke:1996he}. The
following values for the different neutrino temperatures are used,
giving the following average energies:
\begin{center}
\begin{tabular}{llcl} 
$\nu_e$ & $T=3.5$ MeV & $\Rightarrow$ & $<E> = 11$ MeV \\
$\bar\nu_e$ & $T=5.0$ MeV & $\Rightarrow$ & $<E> = 16$ MeV \\
$\nu_{\mu,\tau}$ & $T=8.0$ MeV & $\Rightarrow$ & $<E> = 25$ MeV \\
$\bar\nu_{\mu,\tau}$ & $T=8.0$ MeV & $\Rightarrow$ & $<E> = 25$ MeV \\
\end{tabular}
\end{center}

\begin{figure}[htbp]
\centering
\epsfig{file=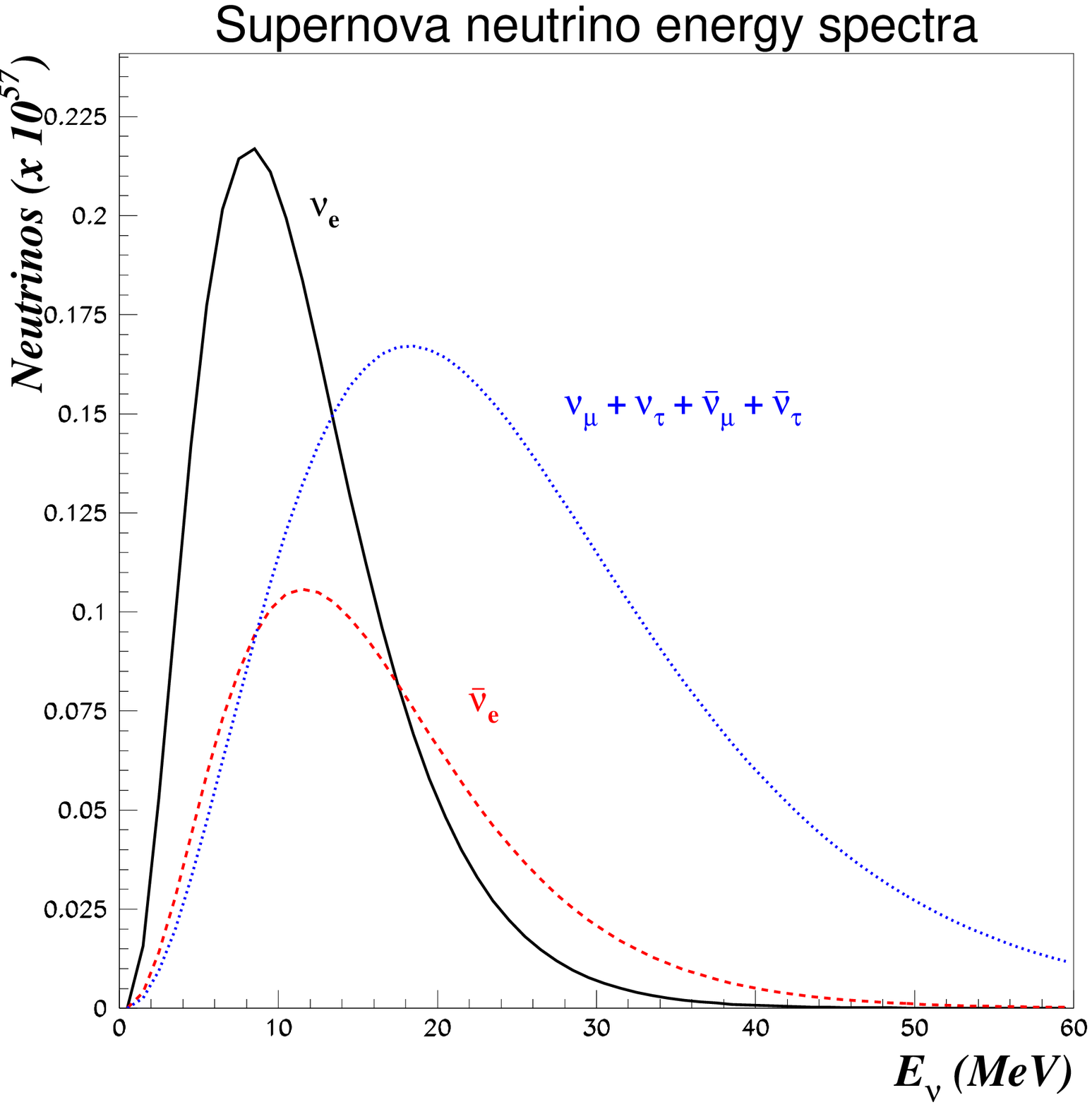,width=15.cm}
\caption{Supernova neutrino energy spectra for a type II supernova 
releasing $3\times 10^{53}$ ergs.}
\label{fig:spec}
\end{figure}

According to~\cite{Qian:1994hh}, pairs of neutrino 
flavours should be produced with equal luminosity. The total number of 
neutrinos of a given flavour is obtained dividing the binding
energy by six (corresponding to all $\nu + \bar\nu$ species) and by the
average energy associated to each species. As a result, $\nu_e$'s
will be more copiously produced than $\nu_{\mu,\tau}$'s since their 
mean energy is
lower. Figure~\ref{fig:spec} shows the expected neutrino spectra for
a binding energy of $3\times 10^{53}$ ergs. We expect
$2.8\times 10^{57} \ \nu_e$, $1.9\times 10^{57} \ \bar\nu_e$ and 
$5.0\times 10^{57} \ \nu_{\mu,\tau}+\bar\nu_{\mu,\tau}$ for a total 
of  $9.7\times 10^{57}$ neutrinos.

In argon two processes contribute to the total rate: 
\begin{itemize}
\item {\bf Elastic scattering:} $\nu_x + e^- \to \nu_x + e^- \ (x=e,\mu,\tau)$ 
sensitive to all neutrino species. 
\item {\bf Absorption:} $\nu_e + ^{40}Ar \to e^- + ^{40}K^*$, 
$\bar\nu_e + ^{40}Ar \to e^+ + ^{40}Cl^*$ 
\end{itemize}

We consider two configurations of the detector with masses
of 1.2 and 3 kton: the number of targets is accordingly:

\begin{tabular}{lclcl} 
& & \hspace*{1.cm}1.2 kton & & \hspace*{1.cm}3 kton \\\cline{3-3}\cline{5-5}
{\bf Elastic scattering } & & $3.2 \times 10^{32} \ e^-$ & 
& $8.0 \times 10^{32} \ e^-$ \\
{\bf Absorption} & & $1.8 \times 10^{31}$ Ar nuclei & & 
$4.5 \times 10^{31}$ Ar nuclei \\
\end{tabular}

\subsection{Elastic events}

The elastic neutrino scattering off electrons has a total cross
section that increases linearly with energy:
\begin{eqnarray}
\begin{tabular}{lccc}
$\sigma(\nu_e e^- \to \nu_e e^- )$ & = & $9.20 \times 10^{-45} E_{\nu_e} {\rm (MeV) \ \ \ \ cm}^2$ \\
$\sigma(\bar\nu_e e^-  \to \bar\nu_e e^- )$ & = & $3.83 \times 10^{-45} E_{\bar\nu_e} {\rm (MeV) \ \ \ \ cm}^2$ \\
$\sigma(\nu_{\mu ,\tau} e^-  \to \nu_{\mu ,\tau} e^- )$ &=& $1.57 \times 10^{-45} E_{\nu_{\mu ,\tau}} {\rm (MeV) \ \ cm}^2$ \\
$\sigma(\bar\nu_{\mu ,\tau} e^-  \to \bar\nu_{\mu ,\tau} e^- )$ &=& $1.29 \times 10^{-45} E_{\bar\nu_{\mu ,\tau}} {\rm (MeV) \ \ cm}^2$ \\
\end{tabular}
\end{eqnarray}

All neutrino species contribute to elastic scattering. The
experimental signature consists in a single recoil electron. Since,
the direction of this electron is highly correlated to the incoming
$\nu$ direction, these events have the potentiality of precisely
determining the location of the supernova source.

Table~\ref{tab:rates} shows the expected rates for this 
reaction. To compute the number of events we fold the total cross
section as a function of energy with the appropriate Fermi-Dirac
distribution. Our rates are calculated integrating over all electron
recoil energies. Since the neutrino burst occurs in a time window of
about 10 seconds, we estimate that the background expected due to
natural radioactivity is negligible and therefore we do not apply 
any threshold for electron detection. 

\begin{figure}[htbp]
\centering
\vspace{-1cm}
\epsfig{file=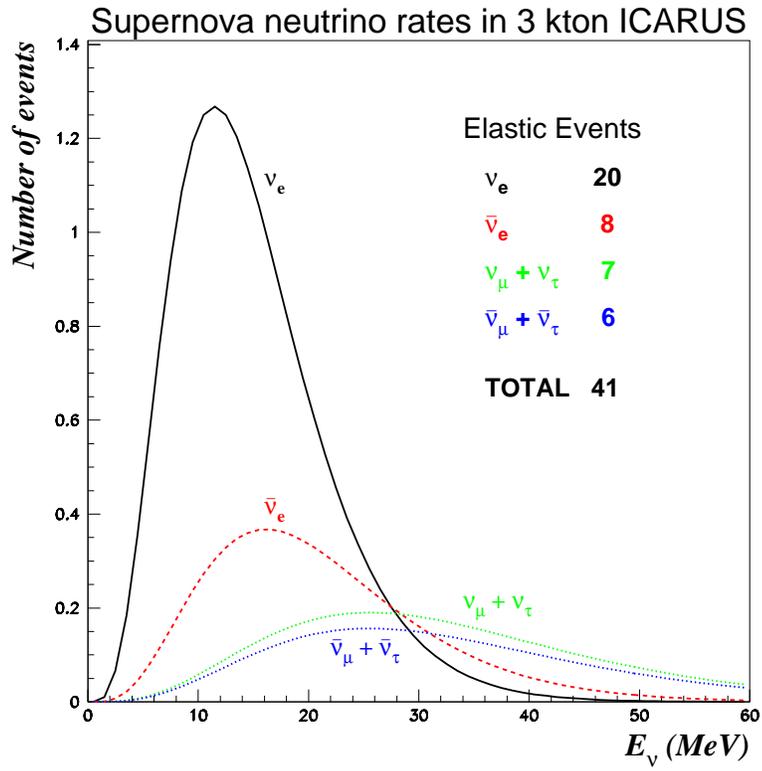,width=11.cm}
\epsfig{file=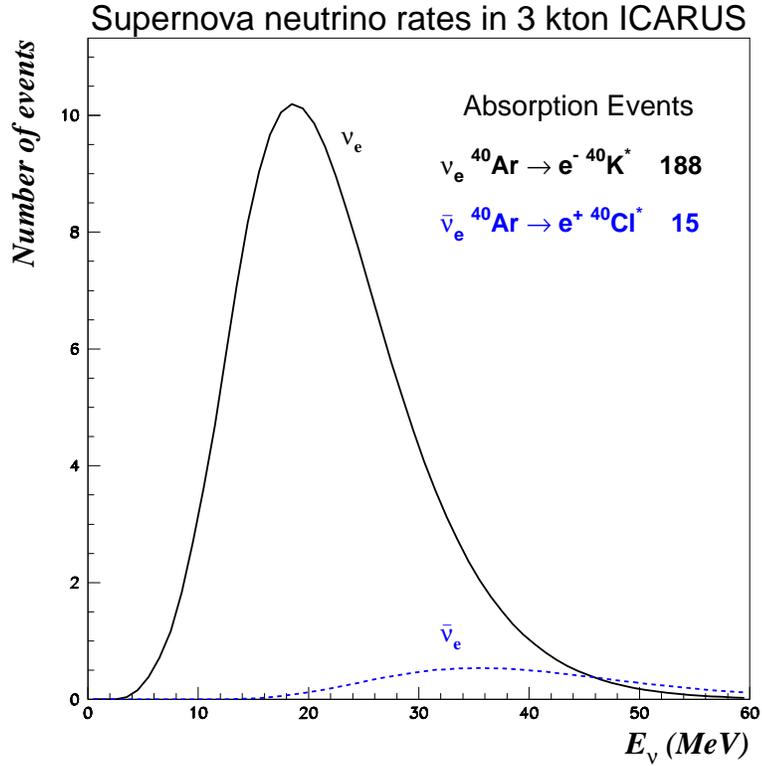,width=11.cm}
\caption{Elastic (top) and absorption (bottom) neutrino event energy
distribution for a type II supernova at a distance of 10 kpc.} 
\label{fig:rates}
\end{figure}

Figure~\ref{fig:rates} shows the expected elastic event rates,
as a function of the incoming neutrino
energy, for the case of a 3 kton detector and a supernova occurring at 
10 kpc. The largest contribution ($\sim 50\%$) to elastic events 
comes from $\nu_e$, since they have both the larger flux and cross
section. For a 3 (1.2) kton detector, we expect a total of 41 (16) elastic
events. 

\begin{table}[htbp]
\centering
\begin{tabular}{clcccc} \hline
& & & &\multicolumn{2}{c}{Expected events} \\
Reaction & & T (MeV) & $<E_\nu>$ (MeV) & 1.2 ktons & 3 ktons \\ \hline
{\bf Elastic} & & & & & \\
& $\nu_e \, e^-$ & 3.5 & 11 & 8 & 20\\
& $\bar\nu_e\, e^-$ & 5 & 16 & 3 & 8\\
& $(\nu_\mu+\nu_\tau)\, e^-$ & 8 & 25 & 3 & 7\\
& $(\bar\nu_\mu +\bar\nu_\tau)\, e^-$ & 8 & 25 & 2 & 6\\
& total $\nu\, e^-$ & & & 16 & 41 \\
\hline
{\bf Absorption} & & & & & \\
& $\nu_e$ $^{40}$Ar       & 3.5 & 11 & 75 & 188 \\ 
& $\anue$ $^{40}$Ar       & 5   & 16 &  6 &  15\\ 
\hline
{\bf Total} & & & & 97 & 244 \\\hline
\end{tabular}
\caption{Expected neutrino rates for a supernova at a distance of 10
kpc, releasing an energy of $3\times 10^{53}$ ergs (no 
threshold on the electron energy has been applied).}
\label{tab:rates}
\end{table}

\subsubsection{Reconstruction of the incoming neutrino direction}
	
It is possible to use the angular distribution of elastic scattering neutrino
events to identify the direction of a nearby supernova. Ordinarily,
this would be unnecessary due to the optical
emissions. However, if one is interested in studying a galactic
supernova's first hours or days or if the supernova is obscured in the
optical by the dust of the galactic disk, the early warning and crude
angular position provided by the neutrino data would be uniquely
useful. 

One method to achieve this is to use the fact that the electron
produced via an elastic scattering is strongly forward-peaked and it
will point back to the supernova direction. Depending on the type of
the incident neutrino and its energy, one can expect the distributions
of the secondary electrons to be confined to a certain effective
scattering cone.    

In order to estimate our angular efficiency, supernova neutrino
events have been fully simulated using the \FLUKA\ Monte Carlo
package~\cite{flukalisb}. A detailed description of the ICARUS T600
geometry is included in the simulation \cite{geometry}.    
The detector internal volume consisted of two half-modules filled with
liquid argon. A fiducial volume of 3.2 $\times$ 3.0 $\times$ 18.0
m$^3$ has been considered for every half-module, surrounded by a dewar
of stainless steel of 20 tons. An external container of aluminum and
honeycomb acts as thermal insulator. Finally, a 70 cm thick layer of  
polyethylene tubes filled with boric acid has been simulated as a
shield against neutrons.  

Elastic events are characterized by the production of a primary
electron track in the sensitive volume in a direction correlated
with the incoming neutrino direction. The electron undergoes
multiple scattering and Bremsstrahlung which leads to a deviation from
the original direction, also affecting the angular reconstruction accuracy. 

The signal deposited by the track in the detector is discretized in
elementary {\it cells}, which correspond to the readout granularity
of the detector. The detector volume is supposed to be divided in
cells of 3 $\times$ 3 $\times$ 3 mm$^3$. Then, the energy associated
to the track is computed from the charge deposited in every elementary
cell. 

The reconstruction of the electron momentum is performed by using the
position of every cell with respect to the primary vertex weighted
with the corresponding charge deposited in the cell. The algorithm can
be optimized by varying the track length (and the consequent number of
cells) considered in the reconstruction of the electron direction to
avoid the multiple scattering effect.  

Two different optimization methods have been tested in terms of two
variables: the percentage of cells belonging to the electron track to
be used in the reconstruction and the scattering angle of the electron.   

\begin{figure}[htbp]
\centering
\epsfig{file=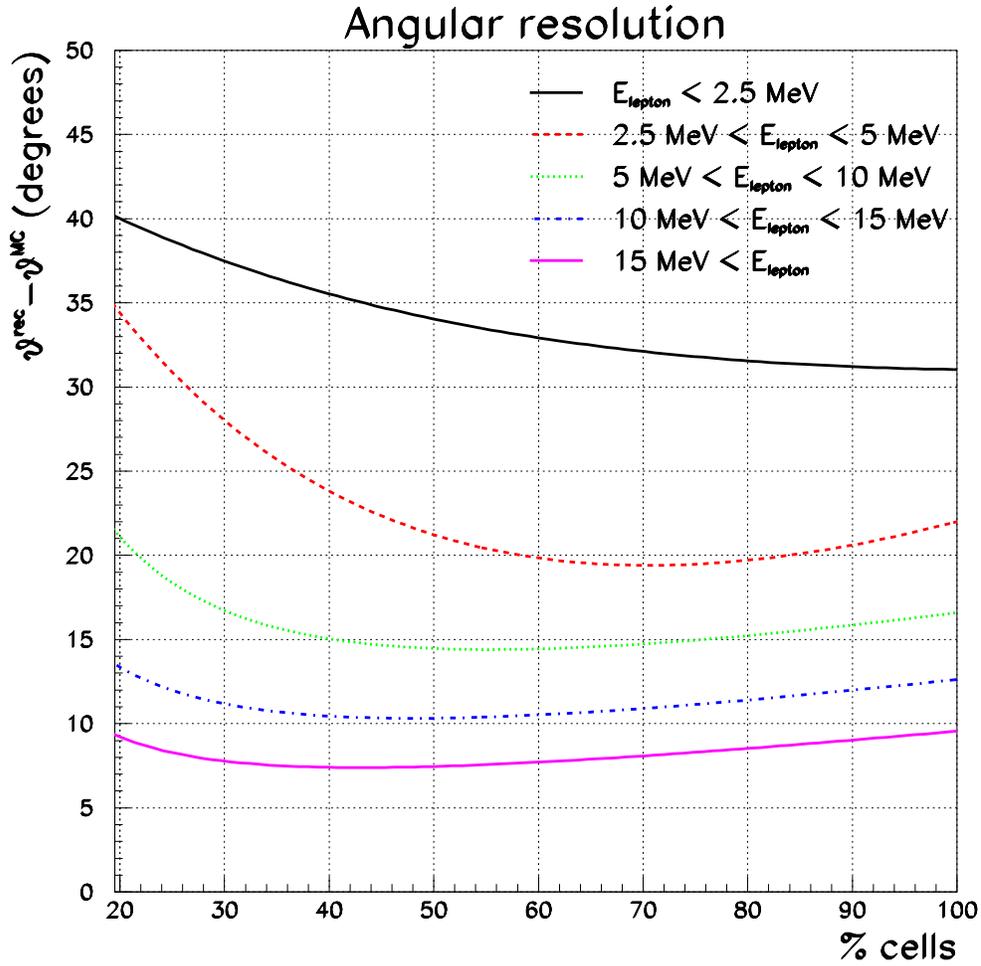,width=14.cm}
\caption{Supernova neutrino angular resolution as a function of the
percentage of cells used in the momentum reconstruction for different
lepton energy ranges.}   
\label{fig:algo1perc}
\end{figure}

In the first method we studied the variation of the angular resolution
as a function of the percentage of cells taking into account in the
momentum reconstruction. The angular resolution was computed as the
RMS of the distribution $\theta^{rec}$--$\theta^{MC}$, the difference
between the reconstructed and simulated angles between the electron
and the parent neutrino directions. Figure \ref{fig:algo1perc} shows
this variation for different electron energy ranges. The percentage of
cells used in the electron direction reconstruction giving the best
angular resolution decreases when the electron energy increases.  

\begin{figure}[htbp]
\centering
\epsfig{file=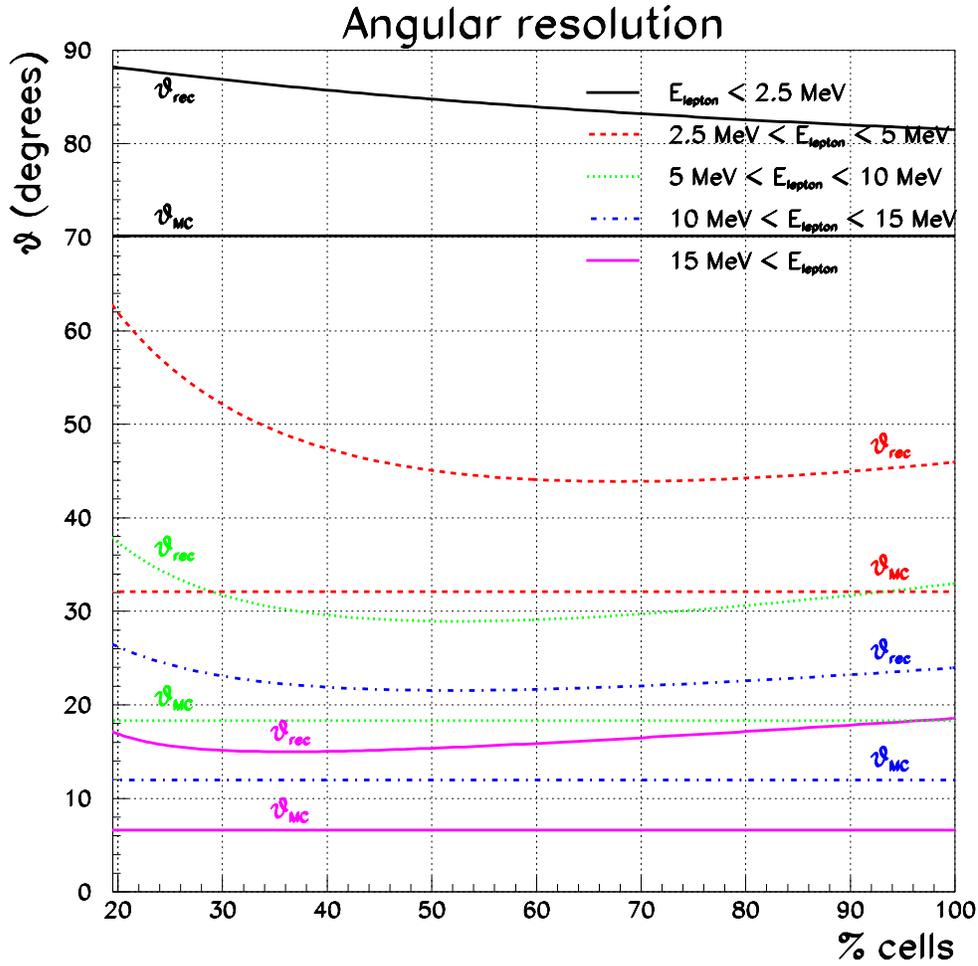,width=14.cm}
\caption{Reconstructed ($\theta^{rec}$) and generated ($\theta^{MC}$)
angles between the leading electron and the incoming neutrino as a
function of the percentage of cells used in the momentum
reconstruction for different lepton energy ranges.}
\label{fig:algo1perc_angs}
\end{figure}

The reconstructed and simulated angles for different electron energy
ranges are plotted separately in figure \ref{fig:algo1perc_angs}. The
value of the angle between the electron and the neutrino is dominated
by the kinematics of the events and not by the reconstruction
algorithm. 

\begin{figure}[htbp]
\centering
\epsfig{file=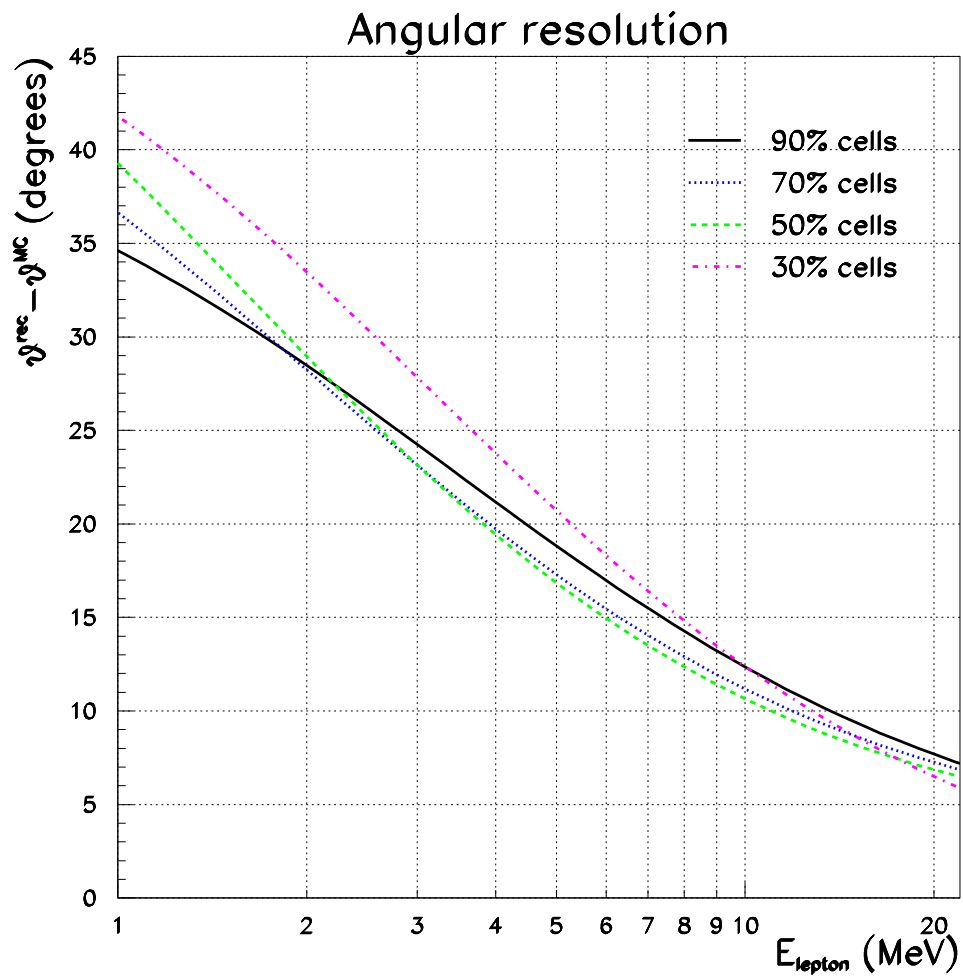,width=14.cm}
\caption{Supernova neutrino angular resolution as a function of the
lepton energy for different percentage of cells used in the momentum
reconstruction.} 
\label{fig:algo1}
\end{figure}

Figure \ref{fig:algo1} shows the expected angular resolution as a
function of the electron energy for different percentage of cells. For
low electron energies ($<$ 2.5 MeV), almost all cells are needed to
have an accurate reconstruction (90\%). For energies between 2.5 and
16 MeV, the optimal resolution is obtained using the 50\% of the cells
and for electrons of more than 16 MeV only the 30\% of the cells
should be used.    

\begin{figure}[htbp]
\centering
\epsfig{file=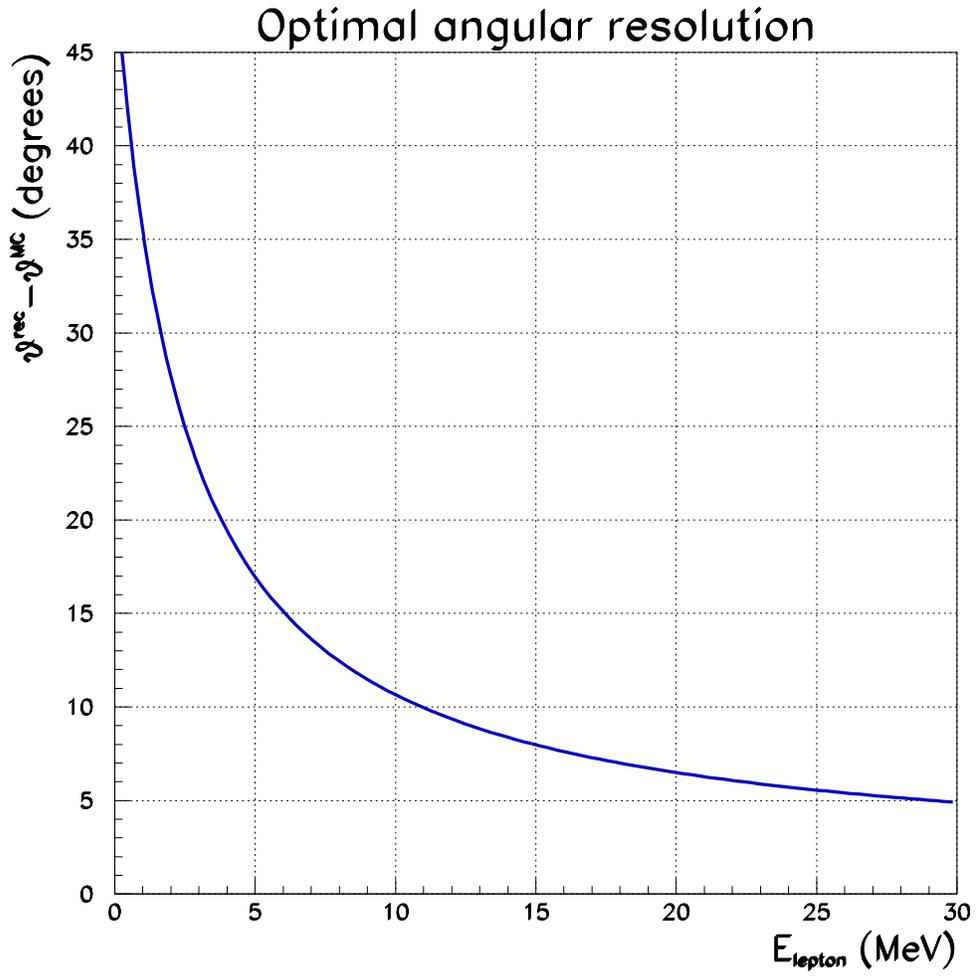,width=14.cm}
\caption{Supernova neutrino angular resolution as a function of the
lepton energy. The optimal algorithm for the momentum reconstruction
is used.} 
\label{fig:algo2}
\end{figure}

Figure \ref{fig:algo2} shows the angular resolution obtained with the
best reconstruction algorithm which uses the optimal percentage of
cells for each electron energy range. The resolution decreases
strongly at low energies, being below 16 degrees for events with
electron energies of more than 5 MeV. For energies greater than 15 MeV
the resolution is less than 8 degrees.

The second method is based on the dependence between the electron
momentum and the scattering angle. The projected scattering angle can
be obtained as \cite{PDG}: 

\begin{equation}
\langle \theta_i \rangle = \frac{13.6~ \textrm{MeV}}{\beta c p} 
\sqrt{\frac{x}{X_0}} \left[1 + 0.038 \ln \left(\frac{x}{X_0}\right) \right]
\label{AVERAGE_ANGLE}
\end{equation}

\noindent where $x$ is the track length, $p$ and $\beta c$ are the
momentum and velocity of the particle, and $X_0$ = 14 cm is the
radiation length in LAr. Therefore the particle momentum, neglecting
ionization losses, is proportional to the inverse of the average value
of the scattering angles.

We select the track length for which the scattering angle is
smaller that a certain threshold. This way, we reduce the effect of
multiple scattering on the angular resolution.

\begin{figure}[htbp]
\centering
\epsfig{file=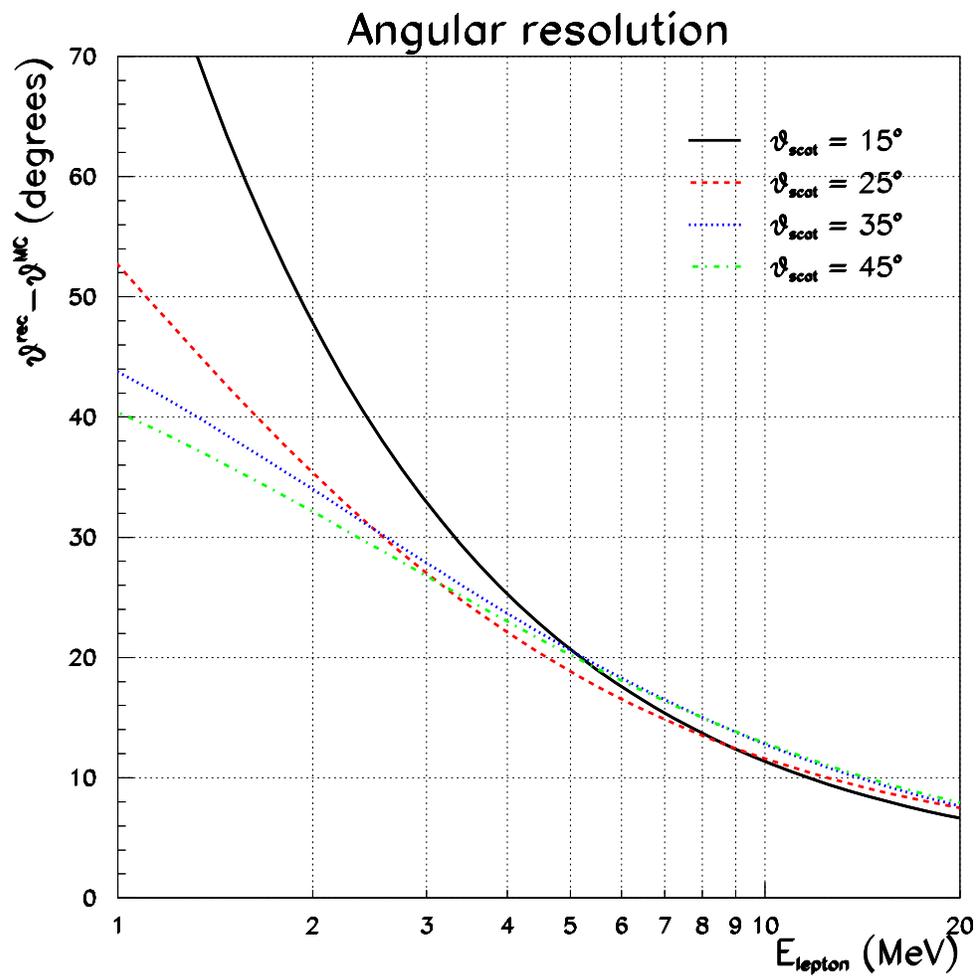,width=14.cm}
\caption{Supernova neutrino angular resolution as a function of the
lepton energy for different values of the scattering angle threshold.}
\label{fig:scat}
\end{figure}

Figure \ref{fig:scat} shows the angular resolution as a function of
the electron energy for different values of the scattering angle
threshold. The same behaviour of the angular resolution as a function
of the electron energy is observed with this method but slightly worse
results are obtained. Events with electrons of more than 5 MeV have an angular
resolution below 18 degrees. For more than 15 MeV, the resolution
reached is essentially the same with both methods. 

Therefore, the first algorithm based in the percentage of cells has
been adopted for the reconstruction of the scattered lepton direction.

\begin{figure}[htbp]
\centering
\epsfig{file=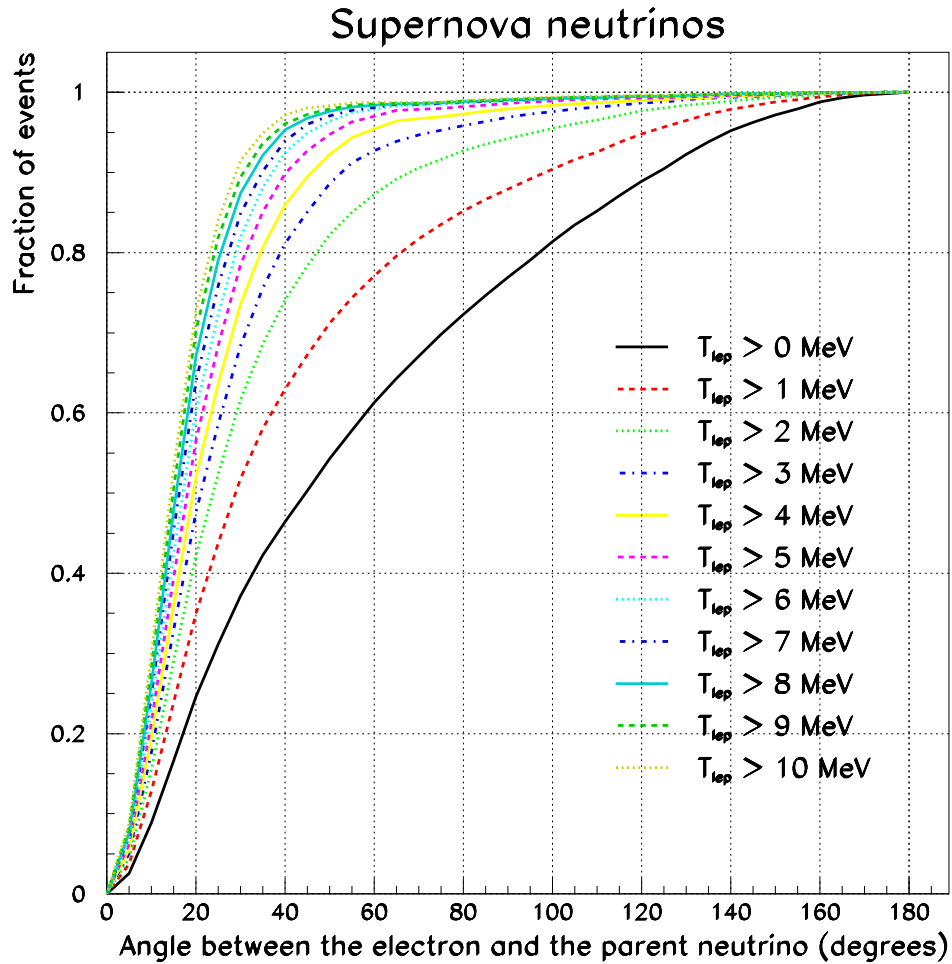,width=14.cm}
\caption{Fraction of events as a function of the angle between the
scattered electron and the parent neutrino direction. The different
curves correspond to different lepton energy thresholds.}  
\label{fig:resol_tlep}
\end{figure}

Figure \ref{fig:resol_tlep} shows the fraction of events as a function
of the cone angle within which the reconstructed electron direction in
space is contained. The cone axis is defined by the parent neutrino 
direction. The different curves correspond to different thresholds on
the electron kinetic energy (T$_{lep}$). For a 25$^\circ$ cone around the neutrino
direction and electron energies greater than 2 MeV, the angular
efficiency is about 51\%. It increases as a function of the threshold
on the electron energy, being 66\% for electrons of more than 5
MeV.         

\begin{figure}[htbp]
\centering
\begin{tabular}{cc}
\epsfig{file=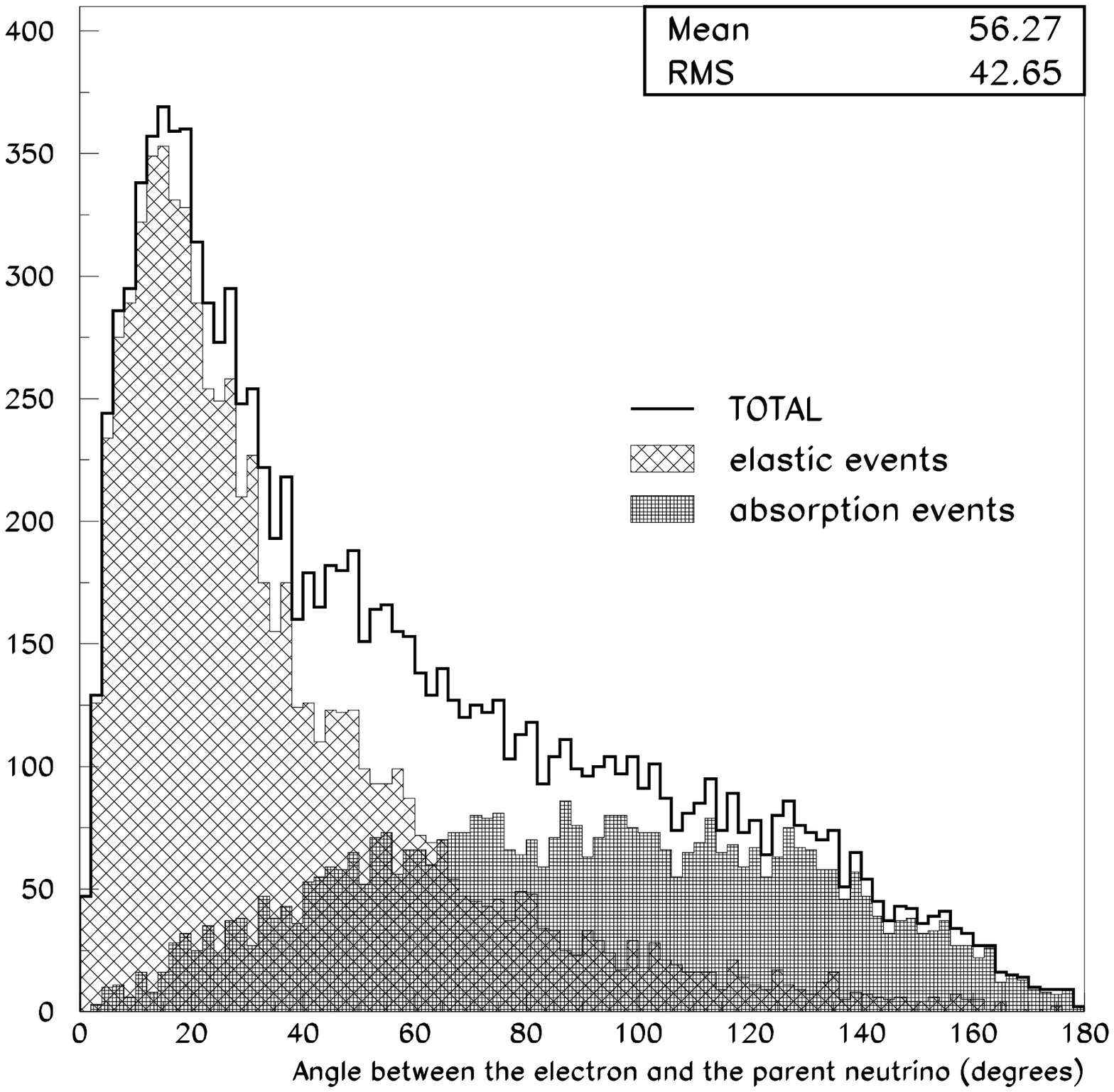,width=8.cm}
&
\epsfig{file=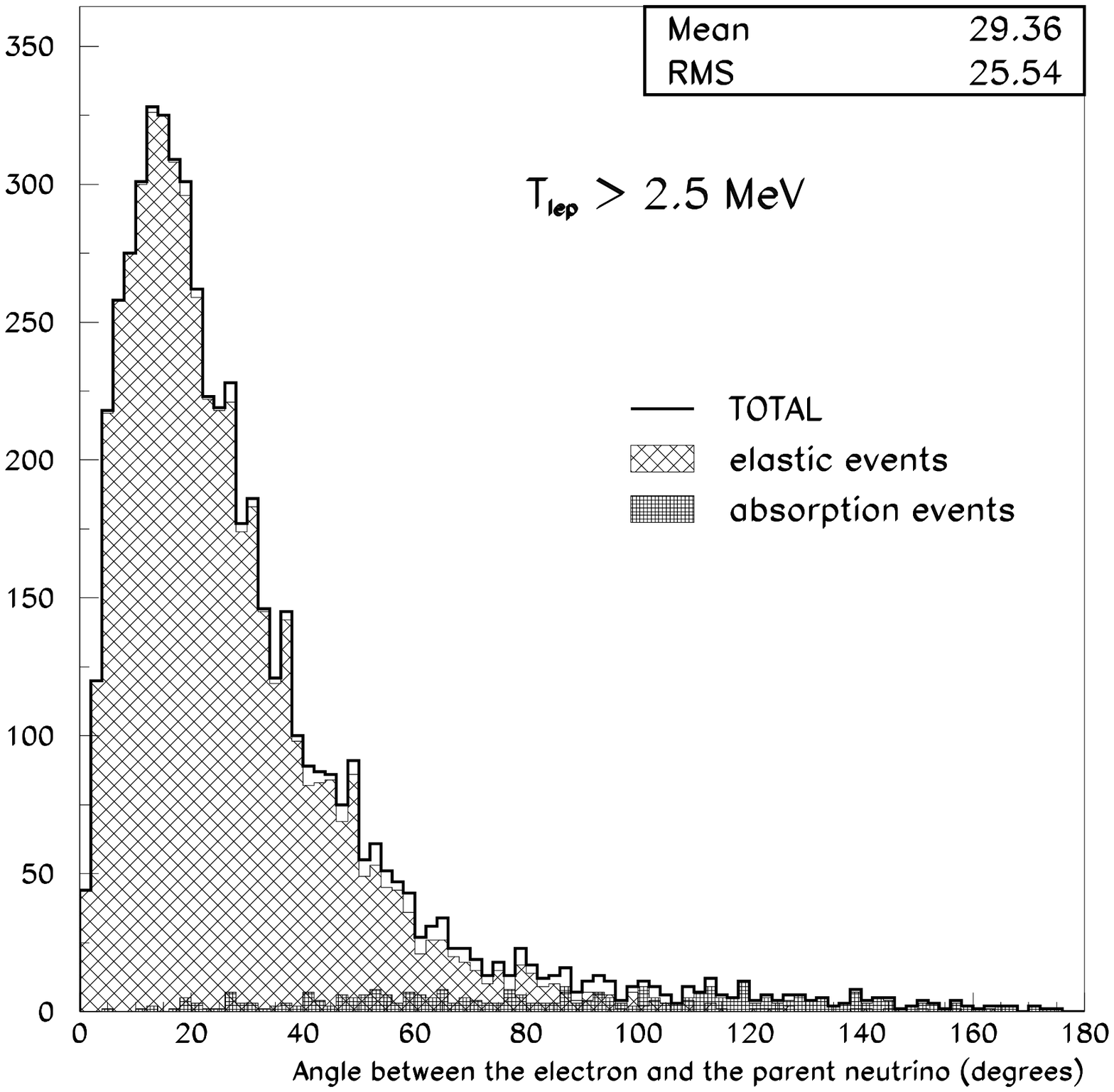,width=8.cm}
\end{tabular}
\begin{tabular}{c}
\epsfig{file=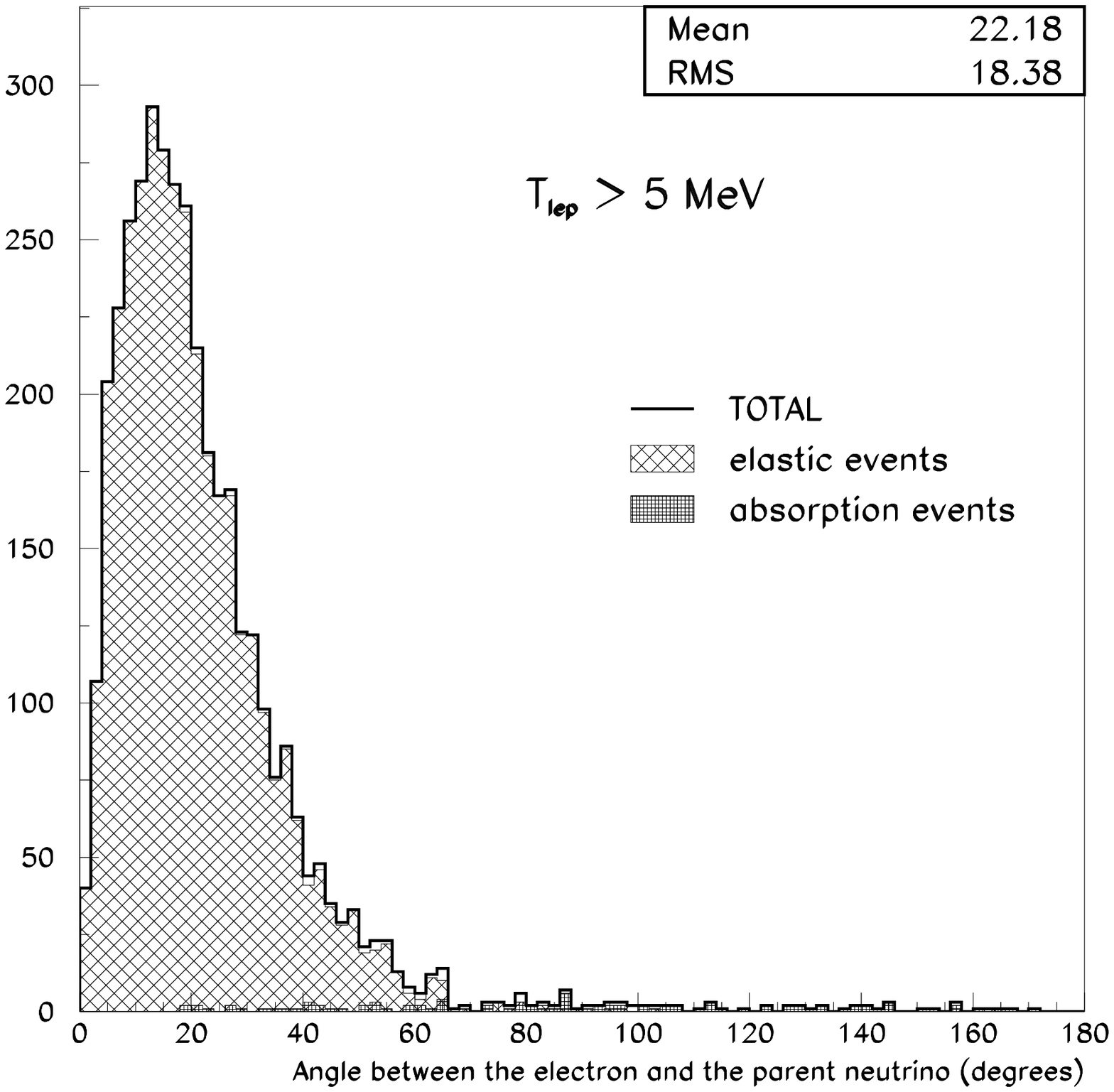,width=8.cm}
\end{tabular}
\caption{Angle between the incoming neutrino and the outcoming
electron for elastic events with electron energies greater than 0, 2.5
and 5 MeV, respectively.} 
\label{fig:ang_eleneu}
\end{figure}

\begin{figure}[htbp]
\centering
\begin{tabular}{cc}
\epsfig{file=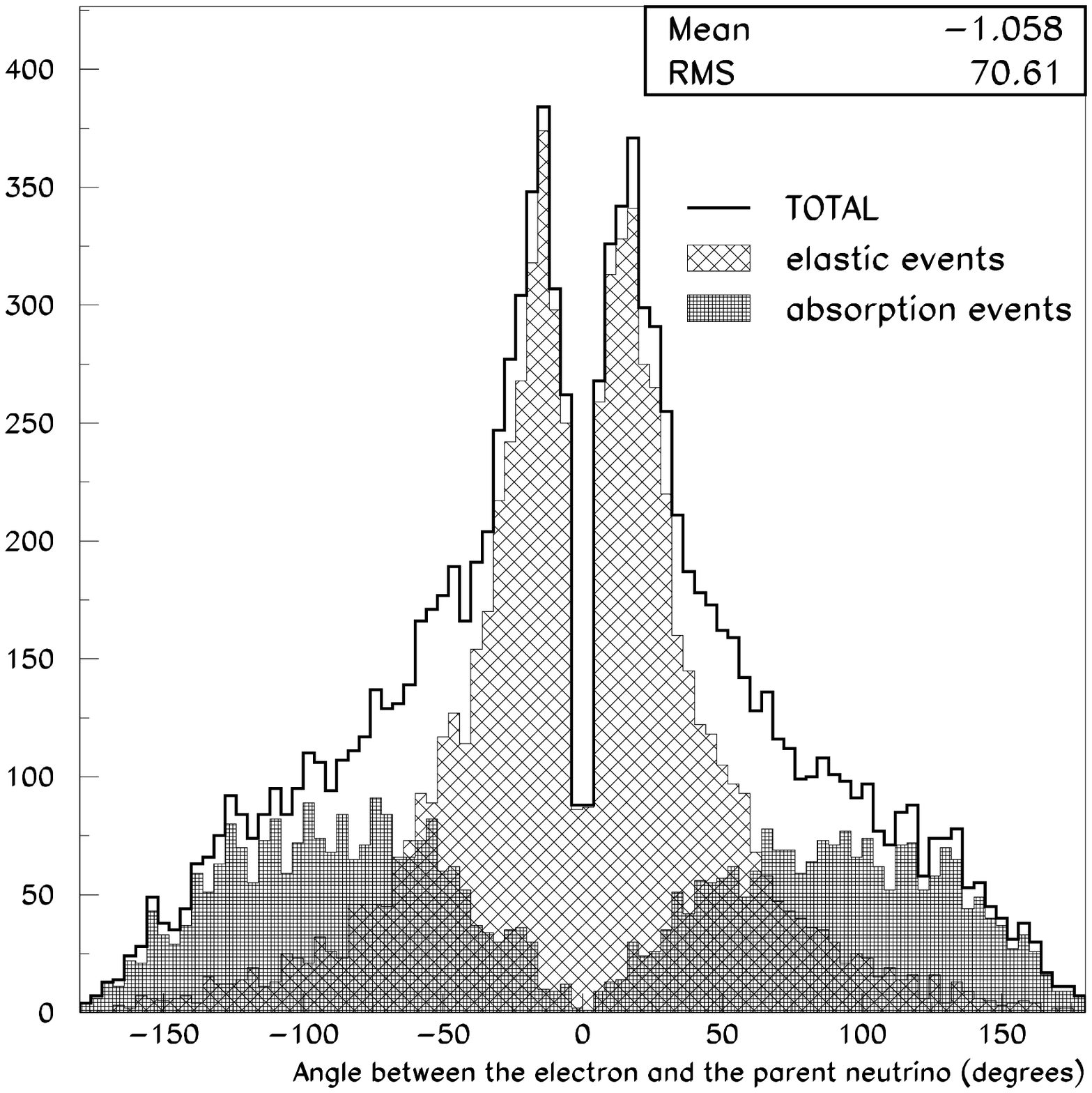,width=8.cm}
&
\epsfig{file=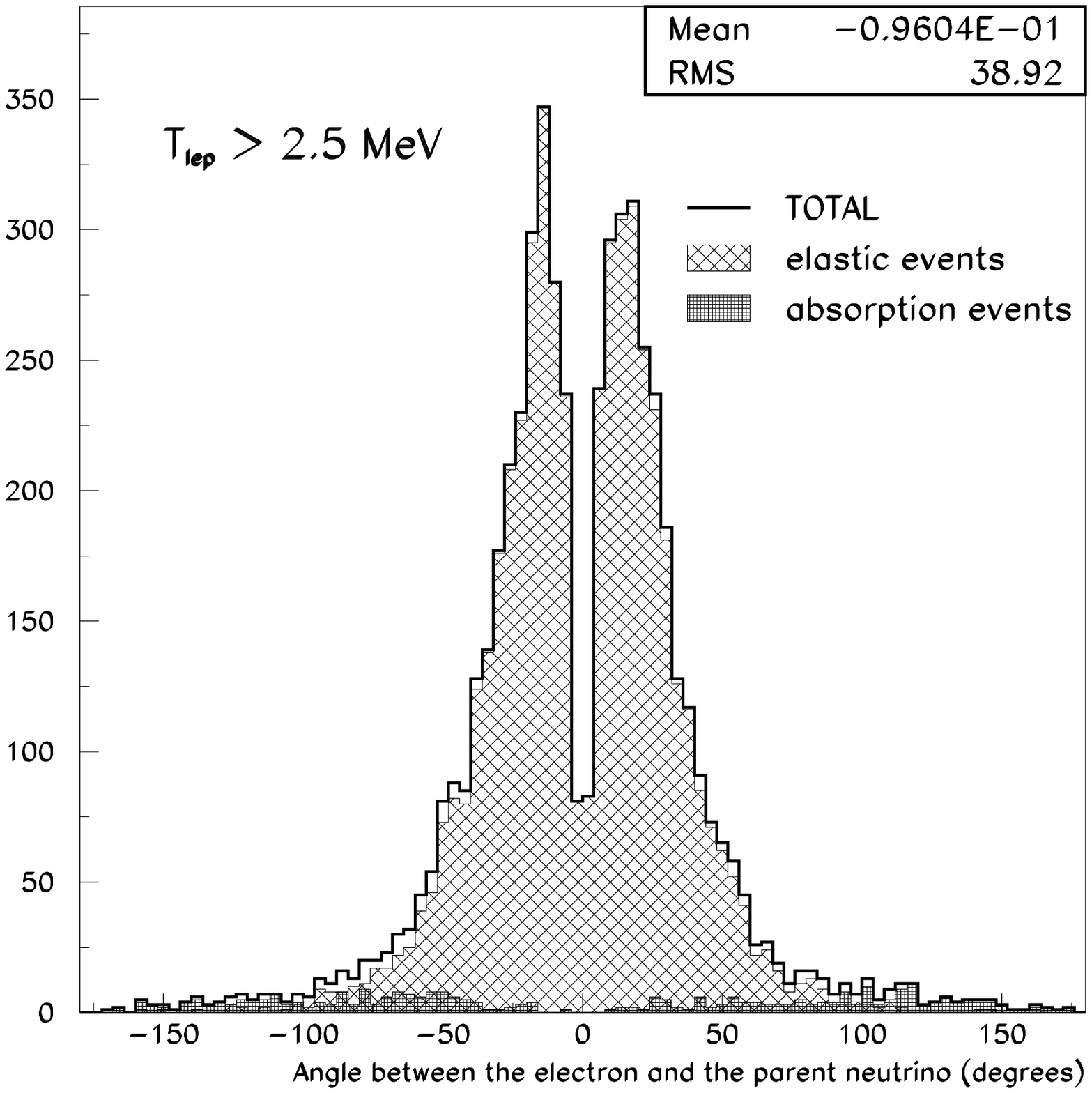,width=8.cm}
\end{tabular}
\begin{tabular}{c}
\epsfig{file=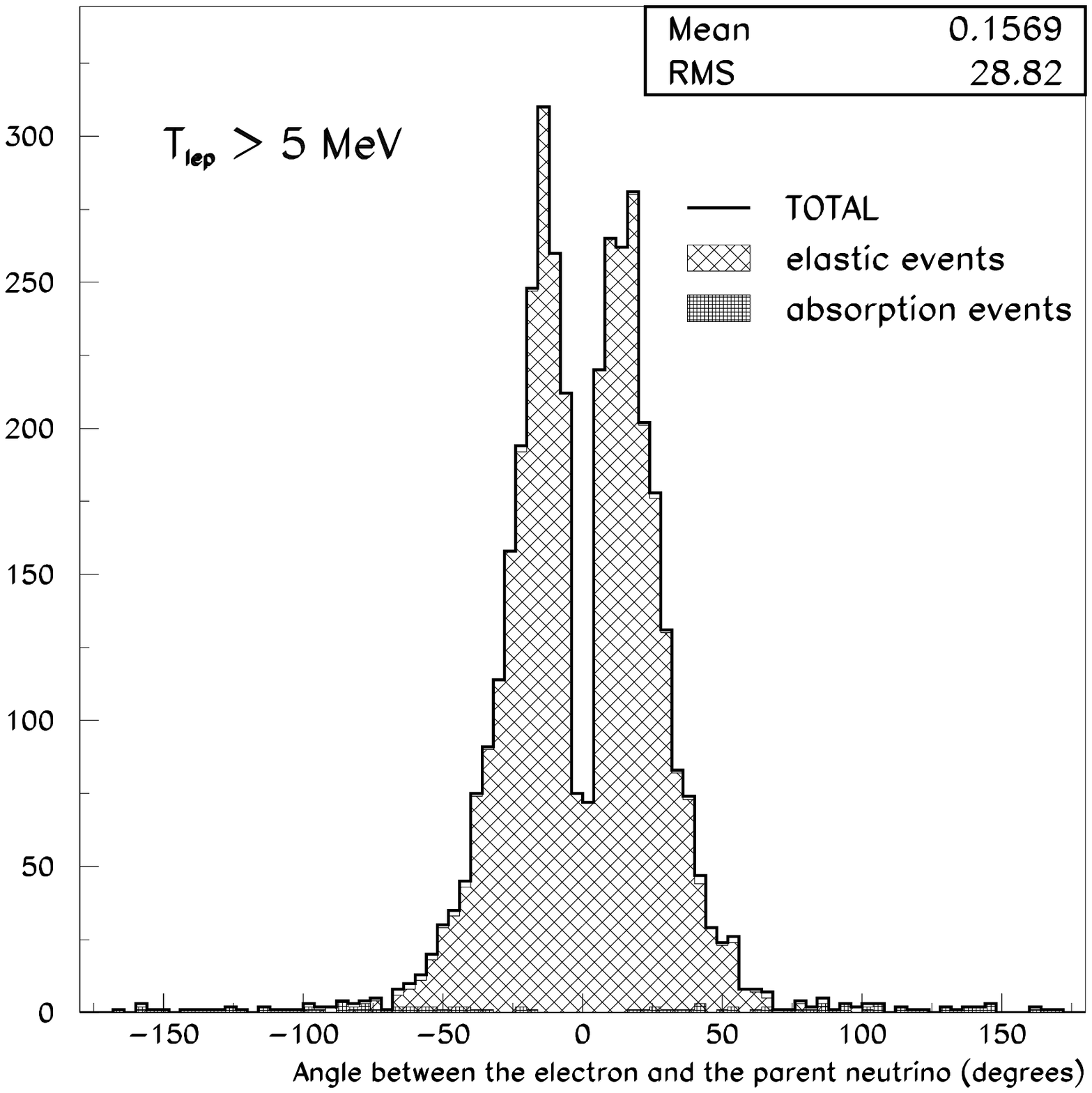,width=8.cm}
\end{tabular}
\caption{Angle between the incoming neutrino and the outcoming
electron for elastic events with electron energies greater than 0, 2.5
and 5 MeV, respectively. The sign of the angle refers to the
projection of the electron momentum into an axis perpendicular to the
neutrino direction.}  
\label{fig:angsig_eleneu}
\end{figure}

In order to determine the supernova explosion location, the angular
distribution of the elastic scattering neutrino events was
used. Figure \ref{fig:ang_eleneu} shows the distribution of the angle 
between the incoming neutrino and the outcoming electron for different
electron energy thresholds. The uncertainty in the direction can be
computed from the RMS of the distributions shown in figure
\ref{fig:angsig_eleneu}. The sign of the angle refers to the
projection of the electron momentum into an axis perpendicular to the
neutrino direction. The contamination coming from absorption events
reconstructed as elastic events is indicated in the plot. It is
largely reduced as soon as we apply a cut on the lepton energy. 

Therefore, the resulting uncertainty in the direction measurement of a
supernova at a distance of 10 kpc is $\sim$ 70.6$^\circ$/$\sqrt{N}$ for
T$_{lep}$ $>$ 0 MeV, $\sim$ 38.9$^\circ$/$\sqrt{N}$ for T$_{lep}$ $>$
2.5 MeV and $\sim$ 28.8$^\circ$/$\sqrt{N}$ for T$_{lep}$ $>$ 5 MeV,
being N the expected number of elastic supernova events. The
corresponding uncertainties obtained with a 1.2 kton and 3 kton
detector are given in table \ref{tab:uncert}, together with the
expected elastic rates for different lepton energy thresholds. 

\begin{table}[htbp]
\begin{center}
\begin{tabular}{|c|cc|cc|}
\hline
{\bf T$_{lep}$} & \multicolumn{2}{|c|}{\bf Expected elastic
events} & \multicolumn{2}{|c|}{\bf Expected uncertainty} \\
\cline{2-5} 
{\bf (MeV)} & {\bf 1.2 ktons} & {\bf 3 ktons} & {\bf 1.2 ktons} & {\bf 3 ktons} \\ 
\hline \hline
$>$ 0 & 16 & 40   & 17.7$^\circ$ & 11.2$^\circ$ \\
$>$ 2.5 & 12 & 30 & 11.2$^\circ$ & 7.1$^\circ$ \\
$>$ 5 & 9 & 23    & 9.6$^\circ$ & 6.0$^\circ$ \\
\hline
\end{tabular}
\end{center}
\caption{Expected uncertainty in the direction measurement of a
supernova at a distance of 10 kpc for a 1.2 kton and 3 kton detector
for different electron energy thresholds.}
\label{tab:uncert}
\end{table}

\subsection{Absorption events}

The $\nue ^{40}Ar$ absorption rate is expected to proceed through two main channels:
a superallowed Fermi transition to the 4.38 MeV excited isobaric
analog $^{40}$K$^*$ state; Gamow-Teller transitions to several excited $^{40}$K
states. 
These two processes can be distinguished by the energy and
multiplicity of the $\gamma$ rays emitted in the de-excitation and by
the energy spectrum of the primary electron.
In addition one can consider other processes which become important
energies above 20~MeV, where the final state is not a bound state
anymore and where emission of particles from the nucleus is possible.

The cross section for the Fermi $\nu_e$ capture is given
by~\cite{Raghavan:1986hv}: 
\begin{equation}
\sigma = 1.702 \times 10^{-44} \frac{E_e}{m_e^2} \sqrt{E_e^2-m_e^2} F(E_e) \ \ {\rm cm^2}
\end{equation}
where $F(E_e)$ is the Fermi function, $E_e = E_\nu - Q + m_e$ is the
prompt electron energy, $Q$ is the energy threshold ($Q = 5.885$ MeV) 
and $m_e$ is the electron mass. 

In reference \cite{Ormand:1995js} shell model calculations of $\nue$
absorption cross sections for $^{40}Ar$ are presented by Ormand {\it et al}. 
They show
that GT transitions lead to a significant enhancement of the
absorption rate over that expected from the Fermi transition.

\begin{figure}[htbp]
\centering
\epsfig{file=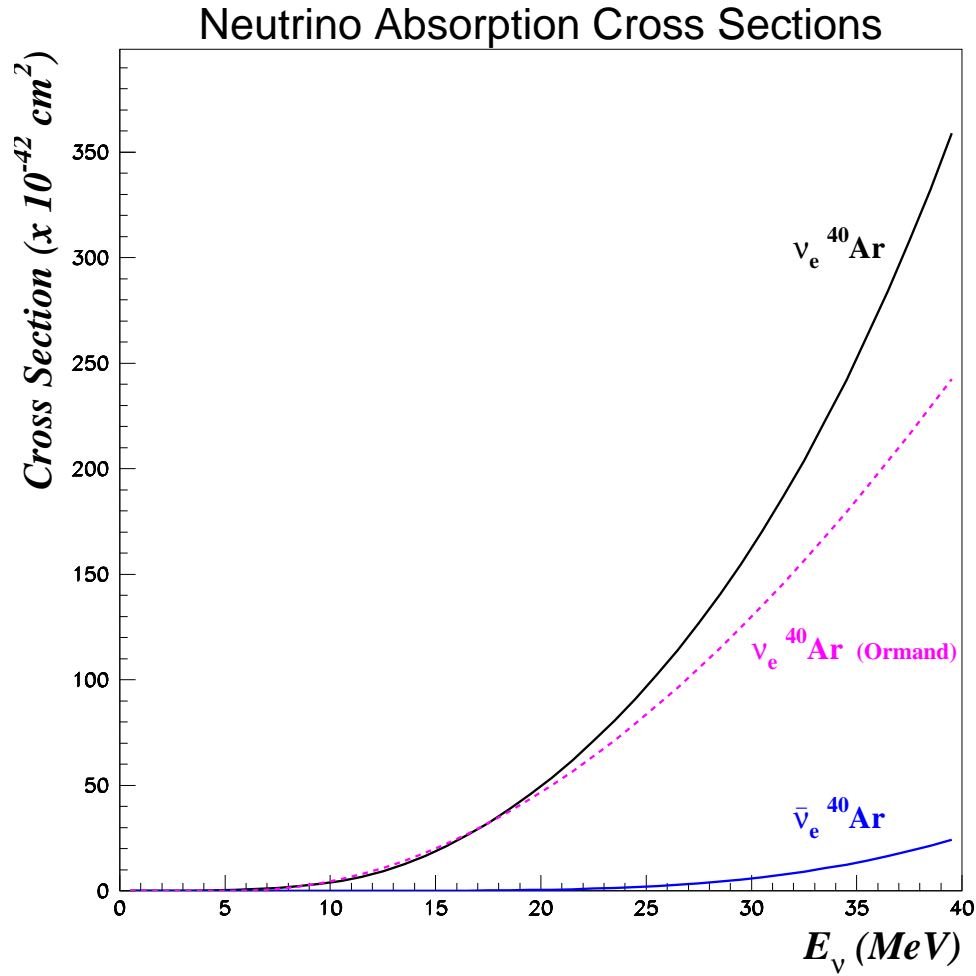,width=14.cm}
\caption{Cross sections as a function of neutrino energy for $\nue$ and
$\anue$ absorption reactions. We compare the results from shell model
calculations \cite{Ormand:1995js} considering only Fermi and
Gamow--Teller transitions with those from RPA calculations including
all the multipoles \cite{martinez}.}    
\label{fig:crossabsorp}
\end{figure}

In recent studies \cite{martinez} the $\nue$ absorption cross section has
been computed by random phase approximation (RPA) for neutrino
energies up to 100 MeV. These calculations include all the multipoles
up to J=6 and both parities, whereas only the 1$^+$ and 0$^+$ allowed
transitions were considered in Ref. \cite{Ormand:1995js}. 
As shown in figure \ref{fig:crossabsorp}, both
results are similar up to $\sim$ 20 MeV. At higher energies relevant for
supernova neutrinos, the cross section including all the multipoles
increases faster, showing the importance of incuding higher
multipoles in order to get a better estimate of the event
rate. 

The absorption cross section is much larger than that for $\nue$
elastic scattering, hence this process significantly enhances the
experimental sensitivity (see table~\ref{tab:rates}). In a 3 kton detector, 
we expect 188 $\nue$ absorption events from the supernova at 10 kpc.  

The absorption process $\bar\nu_e + ^{40}Ar \to e^+ + ^{40}Cl^*$ is
included for the first time, giving 15
events for a 3 kton detector. The corresponding cross section computed
by RPA \cite{martinez} is also plotted in figure \ref{fig:crossabsorp}. In this case
the 1$^+$ and 0$^+$ allowed transitions are suppressed by nuclear
structure reasons and the cross section is dominated by the other
multipoles.  

The total expected rate for a 3 kton (1.2 kton) detector
is about 244 (97) events. This statistics would allow a
detailed study of neutron star and even black hole formation and would
largely contribute to get an improved knowledge of some neutrino
intrinsic properties like, for example, the mass, electric charge and 
magnetic moment.

\begin{figure}[htbp]
\centering
\epsfig{file=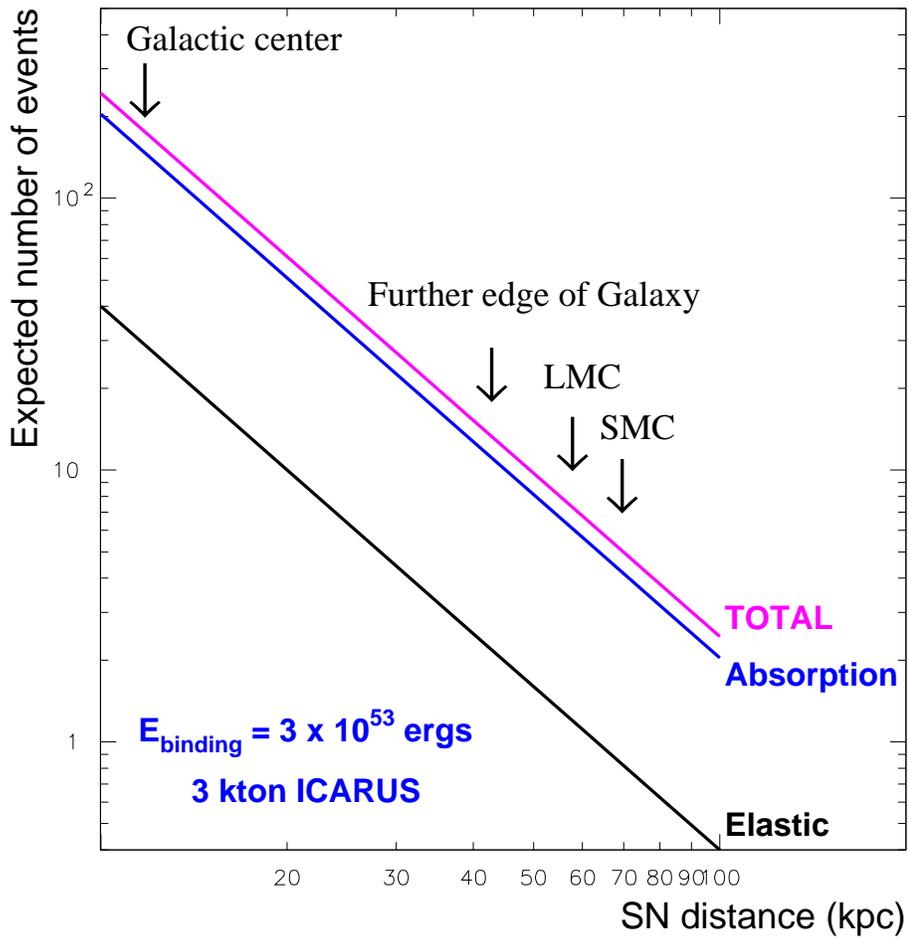,width=14.cm}
\caption{Expected neutrino rates for elastic and absorption reactions 
as a function of the supernova distance. We consider a 3 kton ICARUS-like detector.}
\label{fig:dist}
\end{figure}

Figure~\ref{fig:dist} show the total rate 
for a 3 kton detector as a function of supernova distance. It also
displays the expected rates for the different neutrino reactions. 
For a gravitational stellar collapse occurring in the Large Magellanic
Cloud (distance $\sim$ 60 kpc), we expect to collect, in a 3 kton detector, 
about 7 events as a result of the neutrino burst.

\section{Conclusions}

A liquid Argon TPC like ICARUS is sensitive to supernova neutrinos.
A typical gravitational stellar collapse, occurring at 10 kpc, will
produce a neutrino burst of about 97 (244) events in a 1.2 (3) kton
detector. The largest contribution comes from the
absorption process of $\nu_e$ neutrinos. Elastic scattering off electrons
for all neutrino species contributes to $15\%$ of the total rate. The
$\anue$ absorption reaction has been studied for the first time, giving 15
events for a 3 kton detector. 


The direction of the lepton momentum for elastic scattering
neutrino events can be measured with good accuracy. A
reconstruction algorithm has been optimized by varying the track
length to be considered in the reconstruction. The contamination from
absorption events in the elastic channel can be reduced cutting on the
lepton energy. The expected angular resolution is below 16 degrees for
lepton energies greater than 5 MeV. The resulting uncertainty in the
direction measurement of a supernova at 10 kpc is 17.7$^\circ$
(11.2$^\circ$) with a 1.2 (3) kton detector, reaching
9.6$^\circ$ (6.0$^\circ$) for lepton energies greater than 5 MeV. 

Neutrino oscillations will change significantly the expected number of
events. Matter effects in the SN and vacuum oscillations
will modify the neutrino fluxes. As a result of mixing, the original
$\nue$ spectrum will be shifted to higher energies. This will increase
dramatically the rate of $\nue ^{40}Ar$ absorption events due to
the strong dependence of the cross section with the neutrino
energy. Because of its sensitivity to $\nue$ neutrinos, a liquid Argon
TPC can also provide unique information about the early breakout pulse that
might be observed from the next galactic supernova. Analyses
considering these features will be included in a following paper\cite{new}.

\section*{Acknowledgments}

We would like to acknowledge G. Martinez-Pinedo, E. Kolbe and
K. Langanke for their calculations on the neutrino cross sections in
ICARUS and for very valuable discussions. We also thank A. Ferrari and
P. Sala for their help on the FLUKA simulations.

%
%

%
%

\end{document}